\documentstyle[twocolumn,epsbox]{jpsj2}
\newcommand{\simj}{\stackrel{>}{_\sim}}

\def\runtitle{Superconducting  Transition Temperature of Pr$_{2}$Ba$_{4}$Cu$_{7}$O$_{15-\delta}$}
\def\runauthor{Tatsuro {\sc Habaguchi}, Yoshiaki {\sc \=Ono}, Hai Ying {\sc Du Gh}, Kazuhiro {\sc Sano} and Yuh {\sc Yamada}}

\title
{ Electronic States and Superconducting  Transition Temperature based on 
the Tomonaga-Luttinger liquid in Pr$_{2}$Ba$_{4}$Cu$_{7}$O$_{15-\delta}$ 
}

\author
{
Tatsuro {\sc Habaguchi}, Yoshiaki {\sc \=Ono}, Hai Ying {\sc Du Gh}$^1$, Kazuhiro {\sc Sano}$^1$ and Yuh {\sc Yamada}\\
}
\inst
{
Department of Physics, Niigata University, Ikarashi, Niigata 950-2181, Japan\\
$^1$Department of Physics Engineering, Mie University, Tsu, Mie 514-8507, Japan
}

\recdate
{
\today
}

\abst
{An NQR experiment revealed  superconductivity of Pr$_2$Ba$_4$Cu$_7$O$_{15-\delta}$ (Pr247) to be realized on  CuO double chain layers and suggests  possibility  of novel one-dimensional(1D) superconductivity.
To clarify the nature of the 1D  superconductivity,  we calculate  the band dispersions  of Pr247 by using the generalized gradient approximation(GGA).
It indicates that  Fermi surface of CuO double chains is well described  to the electronic structure of a quasi-1D system. 
 Assuming the zigzag Hubbard chain model to be  an effective model of the system, we derive tight binding  parameters of the model from a fit to the result of GGA. 
Based on the Tomonaga-Luttinger liquid theory, we  estimate  transition temperature ($T_c$) of the quasi-1D zigzag Hubbard  model from the calculated value of the Luttinger liquid parameter $K_{\rho}$.
The result of $T_c$ is  consistent with that of experiments in Pr247
and it suggests that  the mechanism of the superconductivity is well understood  within the concept of the Tomonaga-Luttinger liquid. 
}

\kword
{$\sf{Pr_2Ba_4Cu_7O_{15-\delta}}$, first-principles calculations, 
CuO double chain, superconductivity, Tomonaga-Luttinger liquid
}

\begin{document}
\sloppy
\maketitle

\section{Introduction}
The newly discovered superconductor Pr$_2$Ba$_4$Cu$_7$O$_{15-\delta}$ (Pr247)
 consists of semiconducting single-chain and metallic CuO double-chain besides the Mott insulating CuO$_2$ plane.\cite{Matsukawa,Yamada,Hagiwara}
It shows  the transition temperature ($T_c$)  of superconductivity up to $T_c\sim 20{\rm K}$ in a moderate oxygen defect concentration range of $\delta=0.2 \sim 0.6$, where  $\delta$ can be varied by controlling oxygens in the CuO single chains, and electrons are doped to  CuO double chain.
It is found  that the main carrier is  electron from the measurement of the Hall coefficient at low temperature($T<120$K). 
An NQR (nuclear quadrupole resonance) experiment has confirmed the superconductivity in the CuO double chain layers\cite{Watanabe}.
A spin-gap-like behavior was also observed in the recent NMR experiment\cite{Sasaki}, although it is not clear yet whether the spin gap is  enough large.
These experiments suggest that Pr247 shows the possibility of novel one-dimensional(1D) superconductivity. 

In the previous papers, we address  the 1D double chain models to clarify  the superconductivity of Pr247 from the theoretical point of view.\cite{Sano2005,Sano2007}
Based on the  Tomonaga-Luttinger liquid theory, we  analyze  1D  superconducting mechanism proposed generally by Fabrizio\cite{Fabrizio}, in which the spin gap plays an important role. 
In fact, the spin gap of the double chain model is found to become up to $\sim 100$K  for a typical case and the 1D superconductivity seems to be realistic in Pr247.\cite{Sano2007}
On the other hand,  it is shown that the spin gap  is very small  at some parameter region by using the density matrix renormalization group (DMRG)\cite{Okunishi}. 
Then, Nakano {\it et al.} proposed a spin fluctuation mechanism of the superconductivity for  Pr247 by applying the fluctuation exchange (FLEX) method to
a quasi (1D) extended Hubbard model.\cite{Nakano}
Berg {\it et al.} claimed that "C$_1$S$_{3/2}$" phase, in which the total spin mode is gapless but the  half of the relative spin mode is gaped, is  crucial to  the superconductivity by using the bosonization theory.\cite{Berg}
However, these theoretical works  employed  the result of energy band calculation\cite{Draxl} for not  Pr247 but YBa$_2$Cu$_4$O$_{8}$(Y124) in which CuO double chains are included with the same lattice structure, because the band calculation results of Pr247 has  not been available.
Although  energy dispersions corresponding to the CuO double chain for the both compounds are expected to be almost equivalent, it is very difficult to  avoid  ambiguity of the model to be solved.

In this article,  we calculate the band structure of Pr247 by the density functional calculation with the generalized gradient approximation of
Perdew, Burke, and Ernzerhof by using the WIEN2k package.\cite{wien}
We set the 1D zigzag chain Hubbard model to be an effective model of  CuO double-chain  and obtain the tight-binding parameters of the model by  fitting the energy dispersion.
Using these parameters, we  estimate $T_c$ of the model  based on the Tomonaga-Luttinger liquid theory.
It would give  an accurate analysis of the electronic state of  Pr247 and shed light on the mechanism of the novel quasi-1D superconductivity. 

\section{Band Calculation}
The calculations is done using the WIEN code\cite{wien}, which  uses a full potential-linearized augmented plane wave (FLAPW) calculation. 
 The exchange correlation potential is treated by the generalized gradient approximation of Perdew {\it et al.}\cite{Perdew}. The energy threshold to separate
core and valence states is  -6 Ry. For the number of plane waves the criterion $R_{\rm MT}$ (muffin tin radius) $\times$ $K_{\rm max}$ (for the plane waves) $=7$.
 Self-consistency of the calculation is yield by using a set of 216 k points in the irreducible wedge of the Brillouin zone and the spin-orbit coupling is also considered in our calculations.
The  crystal structure is referred by the experiment values obtained by Yamada {\it et al.}\cite{Yamada}. 
To check the validity of our calculation, we also obtain the result of GGA for Pr123(data not shown) and confirm that it agrees with previous study well\cite{Singh}.

\begin{figure}[t]
\begin{center}
\includegraphics[height=7.5cm]{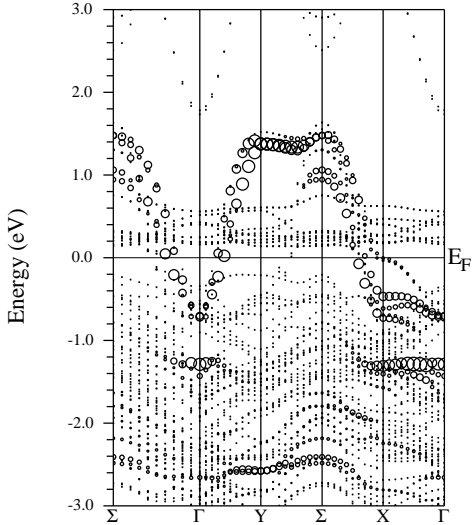}
\end{center}
\caption{A band structure of Pr247 in $k_z=0$ from non-spin polarized calculation. The $\Gamma$ point is at the corners, the $\Sigma$ point in center, the $X$ point at the midpoint of the horizontal edge,
and the $Y$ point at the midpoint of the vertical edge,
}
\label{band}
\end{figure}

Figure \ref{band} shows the band structure of Pr247, where diameter of circles represents weight of contribution  from $d_{x^2-y^2}$ orbitals of Cu in the double chain, where  the double chain is set to be  along the $b$-axis in our calculation.
 We find that seven bands cross  the Fermi energy $E_F$ on $\Sigma-\Gamma$ line among them, two bands which have significant dispersion on $\Gamma-Y$ line,  correspond to the CuO double chain.
It surely shows that CuO double chains are metallic and the dispersion along the $c$-axis perpendicular to CuO$_2$ layer is very small as well as that of Pr123. 
Almost flat bands near and above  $E_F$ consists mainly of   Pr's $4f$ orbital.

\begin{figure}[b]
\begin{center}
\includegraphics[height=5.0cm]{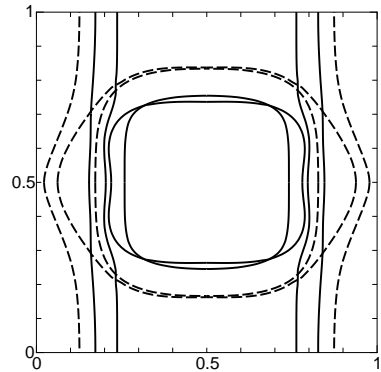}
\caption{A fermi surface in $k_z=0$ for Pr247 from non-spin polarized
calculation.
The $\Gamma$ point is at the corners, the $\Sigma$ point in center, the
$X$ point at the midpoint of the horizontal edge,
and the $Y$ point at the midpoint of the vertical edge,
}
\end{center}
\label{fermi}
\end{figure}

Fermi-surfaces of Pr247  are shown in Fig.\ref{fermi} on the $k_x$-$k_y$ plane with $k_z=0$.
Six vertical lines in both sides present the Fermi-surfaces of single-chain and double-chain, where that of single-chain is denoted by the dashed lines, and the solid lines are corresponding to that of double-chain. 
Four closed Fermi surfaces  are corresponding to $\rm{CuO_2}$ plane.
The hybridization between  double-chain and $\rm{CuO_2}$ plane is very small, 
because the Fermi surfaces of the double-chain are almost straight, though that of the single  chain is warped. 
The result  suggests that the band structure of the double chain is well described as a quasi-1D system. 
Although the existence of the Fermi surfaces of $\rm{CuO_2}$ plane  seems to indicate  $\rm{CuO_2}$ plane to be metallic, it becomes insulator because the electronic conduction in the plane is suppressed due to the so called Fehrenbacher-Rice state formed by the strong hybridization between Pr$4f$ and O$2p$ orbitals\cite{Rice}.
We note that it is  not able to take into account this effect within the GGA calculation. Therefore,  we should not consider the  position of the  Fermi level obtained by GGA  too serious. 

\begin{figure}[t]
\begin{center}
\includegraphics[height=5.0cm]{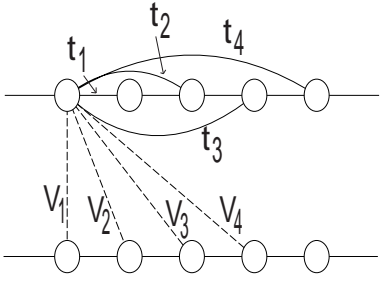}
\end{center}
\caption{The definition of  hopping parameters in the quasi-1D double chain Hubbard model between intra-chain sites and between inter-chain sites, respectively. 
}
\label{model}
\end{figure}

Next, we consider  quasi-1D double chain Hubbard model as a minimum model reflecting the band structure of Pr247. 
It contains  intra-chain hopping integrals between  Cu sites and inter-chain hoppings as shown in Fig. \ref{model} .
The Hamiltonian  is given by
\begin{eqnarray} 
H&=&t_1\sum_{i,m,\sigma}(c_{i,m,\sigma}^{\dagger} c_{i+1,m,\sigma}+h.c.) 
\nonumber \\ 
&+&t_2\sum_{i,m,\sigma}(c_{i,m,\sigma}^{\dagger} c_{i+2,m,\sigma}+h.c.) 
\nonumber \\ 
&+&t_3\sum_{i,m,\sigma}(c_{i,m,\sigma}^{\dagger} c_{i+3,m,\sigma}+h.c.) 
\nonumber \\ 
&+&t_4\sum_{i,m,\sigma}(c_{i,m,\sigma}^{\dagger} c_{i+4,m,\sigma}+h.c.)
+U\sum_{i,m}n_{i,m,\uparrow}n_{i,m,\downarrow} 
\nonumber \\ 
&+&V_1\sum_{i,m,\sigma}(c_{i,m,\sigma}^{\dagger} c_{i,m+1,\sigma}+h.c.) 
 \nonumber \\ 
&+&V_2\sum_{i,m,\sigma}(c_{i,m,\sigma}^{\dagger} c_{i+1,m+1,\sigma}+h.c.)+ 
(c_{i,m,\sigma}^{\dagger} c_{i-1,m+1,\sigma}+h.c.)
 \nonumber \\ 
&+&V_3\sum_{i,m,\sigma}(c_{i,m,\sigma}^{\dagger} c_{i+2,m+1,\sigma}+h.c.) 
+(c_{i,m,\sigma}^{\dagger} c_{i-2,m+1,\sigma}+h.c.)
 \nonumber \\ 
&+&V_4\sum_{i,m,\sigma}(c_{i,m,\sigma}^{\dagger} c_{i+3,m+1,\sigma}+h.c.)+
(c_{i,m,\sigma}^{\dagger} c_{i-3,m+1,\sigma}+h.c.), 
 \label{Hamil}  
\end{eqnarray} 
where $c^{\dagger}_{i,m,\sigma}$ stands for the creation operator of an
 electron with spin $\sigma$ at site $i$ on $m-$th chain and  $n_{i,m,\sigma}=c_{i,m,\sigma}^{\dagger}c_{i,m,\sigma}$.
Here,  $t_{n=1,2,3,4}$ is the hopping energy  between intra chain  sites and $V_{n=1,2,3,4}$   is that  between the nearest-neighbor inter-chain sites as shown in Fig.\ref{model}. 
Here, $U$ is the on-site Coulomb interaction parameter.

In a noninteracting case ($U=0$), the Hamiltonian eq. (\ref{Hamil}) yields 
the band energies as  functions of wave numbers of $k_x$ and $k_y$,
\begin{eqnarray}
E(k_x,k_y)&=&\varepsilon_a + 2t_1\cos (k_x)+2t_2\cos (2k_x)+2t_3\cos (3k_x)
 \nonumber \\ 
&+&2t_4\cos(4k_x)+2V_1\cos(k_y)+4V_2\cos(k_x)\cos(k_y)
 \nonumber \\ 
&+&4V_3\cos(2k_x)\cos(k_y)+4V_4\cos(3k_x)\cos(k_y).
\label{dispersion}  
\end{eqnarray}

\begin{figure}[t]
\begin{center}
\includegraphics[height=7.0cm]{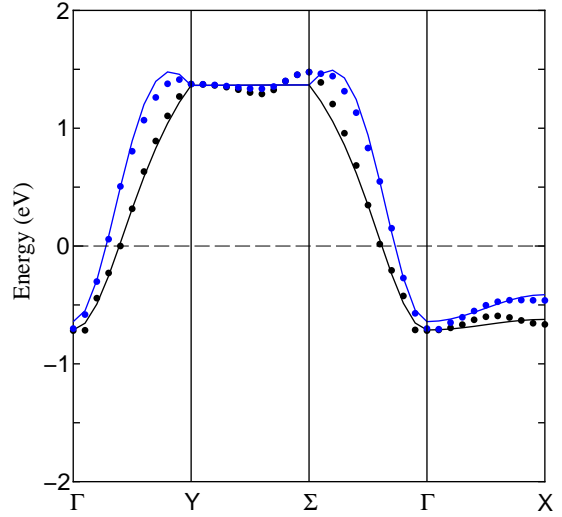}
\end{center}
\caption{The dispersion of the tight-binding model with the parameters determined so as to fit the calculated band structure of Pr24, where the closed circles present GGA result. 
}
\label{fit-band}
\end{figure}

We estimate the tight-binding parameters $t_n$ and $V_n$ of the  model eq.  (\ref{dispersion}) so as to fit  $E(k_x,k_y)$ to the energy dispersion obtained by the GGA.
The obtained  values of parameters are shown in Table I and the fitted dispersion  is shown in Fig.\ref{fit-band}
.
The figure indicates that the obtained dispersion is in good agreement  with the GGA result  and our quasi-1D model is expected to simulate the electronic state of Pr247 very well. 
We also performed same calculation  for Pr124(data not shown) and found that $t_1/t_2$ of Pr124 is smaller than that of Pr247.\cite{C-interaction}
\begin{table}[t]
\begin{tabular}{ccc}
	\hline
	$\hspace{2.5pc}$i$\hspace{2.5pc}$&$\hspace{2.5pc}t_i\hspace{2.5pc}$&$\hspace{2.5pc}V_i\hspace{2.5pc}$\\ \hline\hline
	1&$-0.1227$ eV&$-0.0203$ eV\\ \hline
	2&$-0.4908$ eV&\hspace{0.5pc}$0.0060$ eV\\ \hline
	3&\hspace{0.5pc}$0.0876$ eV&$-0.0097$ eV\\ \hline
	4&$-0.0652$ eV&\hspace{0.5pc}$0.0026$ eV\\ \hline
\end{tabular}
\caption{The values of intra-chain transfer $t_i$, and inter-chain transfer $V_i$ for i-th neighbor sight.}
\end{table}

\section{Tomonaga-Luttinger liquid approach }
Next, assuming the spin gap to be large  enough to exceed the  transition temperature ($T_c$) of the superconductivity\cite{Sano2007,Nishimoto}, we consider $T_c$  by using the formulation of the
Tomonaga-Luttinger liquid in cooperation with the mean-field analysis for quasi-1D systems.\cite{Shultz} 
It contains the  Luttinger liquid parameter $K_{\rho}$ of chain part and  coupling  parameter  $\lambda$ between chains, as shown below
\begin{eqnarray}
T_c = \frac{v_F}{2\pi\alpha}\left\{\frac{\lambda}{4\pi}B^2\left(
\frac{\gamma}{2},\frac{\gamma}{2}\right) \tan{\frac{\pi\gamma}{2}}\right\}
^{1/(2-2\gamma)},	
\label{kr-eq}
\end{eqnarray}
where $\lambda=\frac{4\alpha^2}{{v_F}^2}(2V_1^2+4V_2^2+4V_3^2+4V_4^2$),  $B(x,y)=\frac{\Gamma(x)\Gamma(y)}{\Gamma(x+y)}$, and $\gamma=\frac{1}{2K_{\rho}}$.
Here,  $v_{F}$ is the Fermi velocity.
We approximate  $v_{F}\simeq {v_F}^*$, where ${v_F}^*=\frac{v_{F_1}v_{F_2}}{v_{F_1}+v_{F_2}}$ is an effective Fermi velocity constructed by two Fermi velocities;  $v_{F_1}$ and $v_{F_2}$ in the double chain model.\cite{Nishimoto,Sano2000}  Further, $\alpha$ is a short-wavelength cutoff parameter.\cite{Shultz,Voit} This value  related to the spatial extension  of the quasiparticle, however, it is difficult to determine it without  ambiguity.
We should chose $\alpha \simj 2$, since the dimer state on the nearest neighbor sites is considered to composite main part of the quasiparticle.\cite{Doi,Sano1992,Takano}

In the double chain model with  two different  Fermi points, low-energy excitations are given by a single gapless charge mode with a gapped spin mode\cite{Fabrizio,Balentz,Emery,Kuroki,Daul,Daul2}. In this case, the SC  and  CDW correlations of the charge mode  decay as $\sim r^{-\frac{1}{2K_{\rho}}}$ and $\sim \cos[{2(k_{F_2}-k_{F_1}) r}] r^{-2K_{\rho}}$, respectively, while the SDW correlation decays exponentially. Hence, the SC correlation is dominant for  $K_\rho >0.5$, while the CDW correlation is dominant for $K_\rho <0.5$. 
Since the role of inter-chain couplings  is considered to stabilize the SC state at finite temperature, the electronic state in chain part is  crucial to the creation of the SC.
If $K_\rho >0.5$, the SC correlation is dominant in the chain and the total system becomes superconductor.
%
%
\begin{figure}
\begin{center}
\includegraphics[height=7.0cm]{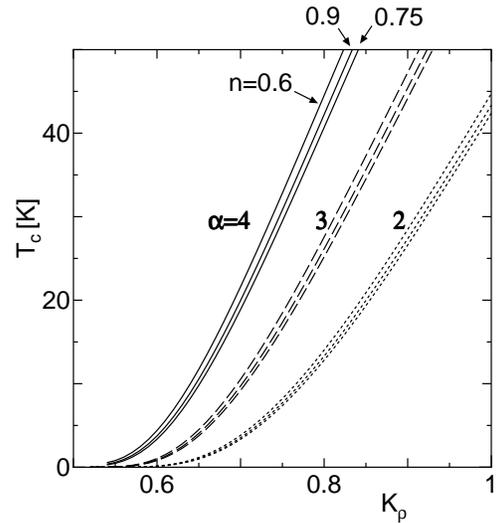}
\end{center}
\caption{$T_c$ as a function of  $K_{\rho}$ at $n=$0.5, 0.75 and 1.0 for $\alpha$=2(dotted line),3(dashed line) and 4(solid line), respectively. 
}
\label{kr-tc}
\end{figure}

In Fig. \ref{kr-tc}, we show  $T_c$  as a function of  $K_{\rho}$ for
severl values of electron number per site $n$ by substituting the fitting parameters of Pr247 into  eq.(\ref{kr-eq}).
At first glance,  the  $n-$dependence of $T_c$ seems to be small, but we stress that  $n-$dependence of $K_{\rho}$ is not small. 
For example, near $n=1$, $K_{\rho}$ becomes  very small due to  the umklapp effect.
 The $n-$dependence  of $K_{\rho}$ for  similar models has been already discussed in several works.\cite{Sano2005,Sano2007,Nishimoto}
They indicate that  $K_{\rho}$ has a maximum value almost around $n\sim 0.7$. 
It  suggests that $T_c$ of Pr247 has also a peak near the same point.

We roughly estimate  $K_{\rho}$ of our model  by the Hartree-Fock approximation; $K_{\rho}=\sqrt{\frac{1}{1+\frac{U}{\pi {v_F}^*}} }$\cite{Nishimoto,Sano2000}, where this approximation is valid only in the weak coupling limit and neglects the umklapp process. 
When $U=2$eV and $n=0.75$, we obtain $K_{\rho} \sim 0.79$ and then $T_c\sim 12$K for $\alpha$=2 and $T_c\sim 23$K for $\alpha$=3.
These results are reasonable and  consistent with the experiment results.\cite{Yamada}
It also  suggests that  the  novel mechanism of the  quasi-1D superconductivity is  realized in Pr247.

\section{Summary and Discussion}

In summary, we calculate the band dispersions  of Pr247 by using the GGA.
It shows that  the electronic state of Pr247 is well described by the  quasi-1D system. 
We also obtain the tight binding  parameters  by fitting the band dispersions to the quasi-1D zigzag Hubbard chain model. 
It would give  an accurate analysis of the electronic state and a concrete basis to address the superconductivity of  Pr247.
On the assumption that the spin gap is large  enough to exceed $T_c$, we estimate $T_c$ of the superconductivity in Pr247 by using the formula of $T_c$ for  quasi-1D superconductor.
The result is  consistent with that of experiments and  suggests that the mechanism of the superconductivity in Pr247 may be understood   within the concept of the Tomonaga-Luttinger liquid.

Finally, we would point out the possibility of the quasi-1D superconductivity in Y124, since  the  double chains are also contained in Y124. 
We estimate $T_c$ of the quasi-1D superconductivity in Y124 as $\sim 40$K  for $U=2$eV and $\alpha$=2  by adopting the tight-binding parameters of Y124 obtained by ref.10,
 where inter-chain hoppings is larger than those in Pr247 resulting higher $T_c$.

At this stage, the experimental evidence of quasi-1D superconductivity has not been explicitly observed, since  the quasi-two-dimensional(2D) superconductivity($T_c\sim 80$K)  in CuO$_2$ plane masks the quasi-1D superconductivity even if it exists.
If we  can suppress only the quasi-2D superconductivity, we expect the quasi-1D superconductivity to be found in Y124  near $40$K. 
In fact, the experiment of Zn-doped Y124 shows that $T_c$  have a plateau at $T_c\sim 25$K in the case of doping rate being larger than 2\%.\cite{Felner} 
If  doped impurities destroy  the superconductivity of CuO$_2$ planes completely, we can expect  that the  plateau may be corresponding to the indication of quasi-1D superconductivity.
Further both of theoretical and experimental studies in this regard is highly desired.




\end{document}